\documentclass[preprint,number,12pt,sort&compress]{elsarticle}

\usepackage{amsmath,amsfonts}
\usepackage{listings}
\usepackage{boxedminipage}
\usepackage{makeidx}
\usepackage{graphicx}

\journal{Computer Physics Communications}

\newcommand{\href}[2]{{\texttt{#2}}}

\newcommand\indexprocedures[1]{\index{procedures!#1}%
	\index{#1|see{procedures}}}
\newcommand\indexsymbols[1]{\index{symbols!#1}%
	\index{#1|see{symbols}}}
\newcommand\indexfunctions[1]{\index{functions!#1}%
	\index{#1|see{functions}}}

\lstdefinelanguage{form}{%
   sensitive=false,
   morecomment=[f][commentstyle][0]{*},
   morecomment=[f][keywordstyle][0]{.},
   morestring=[b]",
   morecomment=[n][stringstyle]{`}{'},
   moredelim=*[directive]\#,
   moredirectives={append,break,call,case,close,commentchar,create,default},
   moredirectives={define,do,else,elseif,enddo,endif,endprocedure,endswitch},
   moredirectives={exchange,external,fromexternal,if,ifdef,ifndef,include},
   moredirectives={message,pipe,preout,procedure,procedureextension,prompt},
   moredirectives={redefine,remove,rmexternal,setexternal,setexternalattr},
   moredirectives={show,switch,system,terminate,toexternal,undefine,write,},
   morekeywords={global, local},
   morekeywords={symmetric, antisymmetric, cyclesymmetric},
   morekeywords={rcyclesymmetric, rcyclic, reversecyclic},
   morekeywords={symbol, symbols, cfunction, cfunctions},
   morekeywords={function, functions, vector, vectors},
   morekeywords={tensor, tensors, ctensor, ctensors},
   morekeywords={set, sets, index, indices, table, ctable},
   morekeywords={dimension, dimensions, unittrace},
   morekeywords={if, else, elseif, endif, while},
   morekeywords={repeat, endrepeat, label, goto},
   morekeywords={argument, endargument, exit},
   morekeywords={inexpression, inside, term},
   morekeywords={endinexpression, endinside, endterm},
   morekeywords={abracket, antibracket, bracket},
   morekeywords={abrackets, also, antibrackets, antisymmetrize},
   morekeywords={argexplode, argimplode, apply, auto, autodeclare},
   morekeywords={brackets, chainin, chainout, chisholm, cleartable},
   morekeywords={collect, commuting, compress, contract},
   morekeywords={cyclesymmetrize, deallocatetable, delete},
   morekeywords={dimension, discard, disorder, drop, factarg, fill},
   morekeywords={fillexpression, fixindex, format, funpowers, hide},
   morekeywords={id, identify, idnew, idold, ifmatch, inparallel},
   morekeywords={insidefirst, keep, load, makeinteger, many, metric},
   morekeywords={moduleoption, modulus, multi, multiply, ndrop},
   morekeywords={nfunctions, nhide, normalize, notinparallel},
   morekeywords={nprint, nskip, ntable, ntensors, nunhide, nwrite},
   morekeywords={off, on, once, only, polyfun, pophide, print},
   morekeywords={printtable, propercount, pushhide, ratio},
   morekeywords={rcyclesymmetrize, redefine, renumber},
   morekeywords={replaceinarg, replaceloop, save, select},
   morekeywords={setexitflag, skip, slavepatchsize, sort, splitarg},
   morekeywords={splitfirstarg, splitlastarg, sum, symmetrize},
   morekeywords={tablebase, testuse, threadbucketsize, totensor},
   morekeywords={tovector, trace4, tracen, tryreplace, unhide},
   morekeywords={unittrace, vectors, write},
   morekeywords={abs_, bernoulli_, binom_, conjg_, count_},
   morekeywords={d_, dd_, delta_, deltap_, denom_, distrib_},
   morekeywords={dum_, dummy_, dummyten_, e_, exp_, fac_},
   morekeywords={factorin_, firstbracket_, g5_, g6_, g7_},
   morekeywords={g_, gcd_, gi_, integer_, invfac_, match_},
   morekeywords={max_, maxpowerof_, min_, minpowerof_},
   morekeywords={mod_, nargs_, nterms_, pattern_, poly_},
   morekeywords={polyadd_, polydiv_, polygcd_, polyintfac_},
   morekeywords={polymul_, polynorm_, polyrem_, polysub_},
   morekeywords={replace_, reverse_, root_, setfun_, sig_},
   morekeywords={sign_, sum_, sump_, table_, tbl_, term_},
   morekeywords={termsin_, termsinbracket_, theta_, thetap_},
   morekeywords={5_, 6_, 7_},
   morekeywords={sqrt_, ln_, sin_, cos_, tan_, asin_, acos_},
   morekeywords={atan_, atan2_, sinh_, cosh_, tanh_, asinh_},
   morekeywords={acosh_, atanh_, li2_, lin_}
}[keywords,strings,comments,directives]

\lstdefinelanguage[spinney]{form}[]{form}{%
   morekeywords=[2]{d,d4,dEps,Gamma5},
   morekeywords=[2]{Spaa,Spab,Spba,Spbb,Spa2,Spb2,SpDenominator},
   morekeywords=[2]{NCContainer,ProjMinus,ProjPlus,Sm,Sm4,SmEps},
   morekeywords=[2]{SpClear,SpCollect,SpContractLeviCivita,SpContractMetrics},
   morekeywords=[2]{SpOpen,SpTrace4,tHooftAlgebra,trL,trR},
   morekeywords=[2]{USpa,USpb,UbarSpa,UbarSpb,SpContract},
   morekeywords=[2]{LightConeDecomposition,LightConeSplitting},
   morekeywords=[2]{RemoveNCContainer,Schouten,ASchouten,BSchouten}
}[keywords,strings,comments,directives]

\lstnewenvironment{form}{%
	\lstset{language=form,%
		frame=single,frameround=tttt,framexleftmargin=6pt,%
		indexstyle=[1]\indexprocedures,%
		indexstyle=[2]\indexsymbols,%
		indexstyle=[3]\indexfunctions,%
		index={[1]SpOpen,SpClose,RemoveNCContainer},%
		index={[3]Spa2,Spb2,Spaa,Spab,Spba,Spbb,Sm,Sm4,SmEps,d,d4,dEps}%
	}%
}{}

\lstnewenvironment{spform}{%
	\lstset{language=[spinney]form,%
		frame=single,frameround=tttt,framexleftmargin=6pt,%
		keywordstyle=[2]{\ttfamily},%
		indexstyle=[1]\indexprocedures,%
		indexstyle=[2]\indexsymbols,%
		indexstyle=[3]\indexfunctions,%
		index={[1]SpOpen,SpClose,RemoveNCContainer},%
		index={[3]Spa2,Spb2,Spaa,Spab,Spba,Spbb,Sm,Sm4,SmEps,d,d4,dEps}%
	}%
}{}

\DeclareMathOperator{\Trace}{tr}
\newcommand{\fmslash}[1]{\ensuremath{/\!\!\!{#1}}}
\newcommand{\pslash}[1][{}]{\fmslash{p}_{#1}}
\newcommand{\qslash}[1][{}]{\fmslash{q}_{#1}}
\newcommand{\kslash}[1][{}]{\fmslash{k}_{#1}}
\newcommand{\tr}[2][{}]{\Trace^{#1}\!\left\{{#2}\right\}}
\newcommand{\FORM}{{\texttt{Form}}}
\newcommand{\spinney}{{\texttt{spinney}}}
\newcommand{\SAM}{\textit{S@M}}
\newcommand{\bra}[1]{\langle #1 \vert}
\newcommand{\brb}[1]{[ #1 \vert}
\newcommand{\kea}[1]{\vert #1 \rangle}
\newcommand{\keb}[1]{\vert #1 ]}
\newcommand{\Spaa}[1]{\langle #1 \rangle}
\newcommand{\Spab}[1]{\langle #1]}
\newcommand{\Spba}[1]{[ #1 \rangle}
\newcommand{\Spbb}[1]{[ #1 ]}
\newcommand{\One}{{\mathbb{I}}}
\newcommand{\Cset}{\mathbb{C}}

\newlength{\funcindent}
\newlength{\funcwidth}
\newenvironment{Ventry}[1]%
 {\begin{list}{}{%
   \settowidth{\labelwidth}{\texttt{#1:}}%
   \setlength{\leftmargin}{\labelsep}%
   \addtolength{\leftmargin}{\labelwidth}}}%
 {\end{list}}

\newenvironment{Procedure}[2]{%
\hspace{.8\funcindent}\begin{boxedminipage}{\funcwidth}
	\raggedright \textbf{\#Procedure} #1(#2)

	\vspace{-1.5ex}

	\rule{\textwidth}{0.5\fboxrule}
	\setlength{\parskip}{2ex}
}{\end{boxedminipage}}

\newenvironment{NFunction}[2]{%
\hspace{.8\funcindent}\begin{boxedminipage}{\funcwidth}
	\raggedright \textbf{NFunction} #1(#2)

	\vspace{-1.5ex}

	\rule{\textwidth}{0.5\fboxrule}
	\setlength{\parskip}{2ex}
}{\end{boxedminipage}}

\newenvironment{CFunction}[2]{%
\hspace{.8\funcindent}\begin{boxedminipage}{\funcwidth}
	\raggedright \textbf{CFunction} #1(#2)

	\vspace{-1.5ex}

	\rule{\textwidth}{0.5\fboxrule}
	\setlength{\parskip}{2ex}
}{\end{boxedminipage}}

\newenvironment{NTensor}[2]{%
\hspace{.8\funcindent}\begin{boxedminipage}{\funcwidth}
	\raggedright \textbf{NTensor} #1(#2)

	\vspace{-1.5ex}

	\rule{\textwidth}{0.5\fboxrule}
	\setlength{\parskip}{2ex}
}{\end{boxedminipage}}

\newenvironment{CTensor}[2]{%
\hspace{.8\funcindent}\begin{boxedminipage}{\funcwidth}
	\raggedright \textbf{CTensor} #1(#2)

	\vspace{-1.5ex}

	\rule{\textwidth}{0.5\fboxrule}
	\setlength{\parskip}{2ex}
}{\end{boxedminipage}}

\newenvironment{Parameters}{%
	\setlength{\parskip}{1ex}
	\textbf{Parameters}
	\vspace{-1ex}
	\begin{quote}}{%
	\end{quote}}

\makeindex

\begin{document}
\begin{frontmatter}
\title{\vspace{-2.5cm}\hfill {\small\rm Nikhef~2010-019}\\ 
\hfill {\small\rm Edinburgh~2010-021}\vspace{1cm}\\
\spinney: A \FORM\ Library for Helicity Spinors}

\author{G.~Cullen}
\address{%
The University of Edinburgh,
School of Physics,
Edinburgh EH9 3JZ,
UK
}
\ead{g.j.cullen@sms.ed.ac.uk}

\author{M.~Koch-Janusz}
\address{Utrecht University, 3508 TC Utrecht, The Netherlands}
\ead{maciekkj@nikhef.nl}

\author{T.~Reiter}
\address{Nikhef, Science Park 105, 1098 XG Amsterdam,
             The Netherlands}
\ead{thomasr@nikhef.nl}
\ead[url]{http://www.nikhef.nl/\~{}thomasr/}

\begin{abstract}
In this work, the library \spinney\ is presented, which provides
an implementation of helicity spinors and related algorithms
for the symbolical manipulation program \FORM. The package is
well suited for symbolic amplitude calculations both in traditional,
Feynman diagram based approaches and unitarity-based techniques.
\end{abstract}

\end{frontmatter}

{\bf PROGRAM SUMMARY}

\begin{small}
\noindent
{\em Manuscript Title:} 
\spinney: A \FORM\ Library for Helicity Spinors\\
{\em Authors:} T.~Reiter\\
{\em Program Title:} \spinney\\
{\em Programming language:} Form 
{\em Classification:}
   4.4 Feynman diagrams,
   5 Computer Algebra,
   11.1  General, High Energy Physics and Computing\\
{\em Nature of problem:} Implementation of the spinor-helicity formalism\\
%
{\em Solution method:} Form implementation\\
\end{small}

\newpage

\hspace{1pc}
{\bf LONG WRITE-UP}

\section{Introduction}
The success of current and future collider experiments depends on a precise
prediction of the expected cross-sections for both signal processes and
their corresponding backgrounds. The experimental precision of these
colliders can only be matched by theoretical calculations beyond leading 
order in perturbation theory. Amongst the most pressing problems are
the virtual (one-loop) corrections to processes with up to four particles
in the final state.
A disection of these one-loop amplitudes into
gauge invariant subamplitudes and the use of a compact notation
allows for the generation of numerically stable, fast
computer programs for their evaluation. The authors of~\cite{Xu:1987}
introduced the \emph{spinor helicity} formalism for massless particles
in their cross-section calculations. Together with its extensions for
massive particles, this approach leads to a compact representation
of helicity amplitudes.

Although recently a number of promising purely numerical methods
for the calculation of one-loop amplitudes have been
presented~\cite{Berger:2008ag,Berger:2009zg,Berger:2009ep,Berger:2010vm,%
Giele:2008bc,Ellis:2008ir,Ellis:2009zw,KeithEllis:2009bu,%
Bevilacqua:2009zn,Bevilacqua:2010ve,%
vanHameren:2009dr,VanHameren:2009vq,Ossola:2006us,Mastrolia:2010nb,
Melia:2010bm},
semi-nu\-me\-ri\-cal and algebraic methods are still an important tool
for matrix element
calculations~\cite{%
Denner:2005nn,Denner:2005fg,Bredenstein:2009aj,Bredenstein:2010rs,%
Binoth:2009rv,%
Britto:2004nc,Britto:2006sj,Mastrolia:2006ki,Forde:2007mi,%
BjerrumBohr:2007vu,Kilgore:2007qr,Badger:2008cm,%
Mastrolia:2009dr,Mastrolia:2009rk,%
Berger:2006sh,Badger:2006us,Badger:2007si,Glover:2008ffa,Badger:2009hw,%
Dixon:2009uk,Badger:2009vh}.
The interested reader finds a more complete list of methods for one-loop
ampitudes and their applications in~\cite{Binoth:2010ra,Bern:2008ef}.

The algebraic manipulations required in matrix element calculations
are very often simple, local operations that have to be applied to millions
of terms, not requiring the knowledge of the whole expressions at
any given moment\footnote{An example for a non-local algorithm is the
factorization of a polynomial, which needs to know all its terms at once.}.
General purpose computer algebra programs easily hit the memory limits
when dealing with expressions of this size as they hold the whole expression
in memory at every point in the program. A different approach is persued
by the symbolic manipulation program \FORM{}~\cite{Vermaseren:2000nd}.
The program provides only
local manipulations of expressions, a fact which is reflected by the
memory model,
viewing an expression merely as a stream of terms; only a single term needs
to reside in memory at any moment while the rest of the expression is stored
on a storage device. In practise, a system of buffers is used to
reduce the number of disk operations and multiple terms are processed at
once on multi-processor systems.

The spinor helicity formalism has been implemented in the computer algebra
system Mathematica through the package \SAM~\cite{Maitre:2007jq}.
It provides routines for the algebraic manipulation and numerical evaluation
of spinor products and Dirac matrices. Mathematica's numerical and algebraic
capabilities paired with a user-friendly interface are a clear advantage
for many calculations. However, the above mentioned memory limitations
restrict its usage to smaller problems. More involved cross-section
calculations require an implementation of spinors in \FORM{}, combining
the elegance of the spinor helicity formalism with \FORM{}'s ability
to process huge expressions. In this article we demonstrate that it is
very easy to exploit \FORM{}'s existing constructs to implement
helicity spinors. Therefore the implementation is in the guise of a
\FORM{} library, \spinney{}, rather than an extension of the core language.
The naming of many of the functions and procedures makes reference to
\SAM{} allowing the user an easier migration between the two libraries.

This article is structured as follows: Section~\ref{sec:theory} gives an
overview over the underlying theory and establishes the notation used
in the rest of this work. Section~\ref{sec:interface} provides an interface
documentation of the provided procedures and the defined symbols and functions.
A couple of examples are discussed in Section~\ref{sec:examples}.

\section{Theoretical Background}\label{sec:theory}

\subsection{Conventions}
We consider spinors $u(p)$ and $v(p)$
which are solutions of the Dirac equations
\begin{equation}
(\fmslash{\hat{p}} - m\cdot\One)u(p)=0\quad\text{and}\quad
(\fmslash{\hat{p}} + m\cdot\One)v(p)=0\text{.}
\end{equation}
Here, we denote $\One$ the identity operator in spinor space and
$\pslash=g_{\mu\nu}\gamma^\mu p^\nu$. Dimension splitting is
understood in the way that for $n\in\Cset$
\begin{subequations}\label{eq:dimension-splitting}
\begin{align}
   g^{\mu\nu}&=\hat{g}^{\mu\nu}+\tilde{g}^{\mu\nu}\\
\intertext{such that}
   \hat{g}^\mu_\mu&\equiv \hat{g}^{\mu\nu}g_{\mu\nu}=4,\\
   \tilde{g}^\mu_\mu&\equiv \tilde{g}^{\mu\nu}g_{\mu\nu}=n-4%
	\quad\text{and}\\
   \hat{g}^{\mu\rho}\tilde{g}_{\rho\nu}&=0\text{.}
\end{align}
\end{subequations}
The projector into the four-dimensional, physical Minkowski space
is of the form
\begin{equation}
\hat{g}=\mathsf{diag}(1,-1,-1,-1)\oplus\mathbf{0}^{(n-4)}\text{,}
\end{equation}
i.e. it leads to the on-shell condition $\hat{p}^2-m^2=0$.

As a short-hand notation we add twiddles and hats to all vector-like
objects to indicate projections into the $4$ and $(n-4)$ dimensional
subspaces ($\tilde{\gamma}^\mu\equiv\tilde{g}^\mu_\nu\gamma^\nu$,
$\hat{k}^\mu\equiv\hat{g}^\mu_\nu k^\nu$, etc.). We work with
anti-commuting $\gamma_5$ in 4 dimensions, which means
\begin{subequations}\label{eq:tHoofAlgebra}
\begin{align}
\{\gamma^\mu,\gamma^\nu\}&=2g^{\mu\nu}\quad\text{and}\\
\{\gamma_5,\hat{\gamma}^\mu\}&=[\gamma_5,\tilde{\gamma}^\mu]=0\text{.}
\end{align}
\end{subequations}
The characteristic equation $(\gamma_5)^2-\One=0$ allows to introduce
the usual projectors $\Pi_L$ and $\Pi_R$ into the left-handed and
right-handed subspace,
\begin{equation}
\Pi_L=\Pi_-=\frac12\left(\One-\gamma_5\right)\quad\text{and}\quad%
\Pi_R=\Pi_+=\frac12\left(\One+\gamma_5\right)\text{.}
\end{equation}
The helicity eigenstates of massless spinors are denoted by a commonly used
bracket notation\footnote{We identify the vectors with their labels where
this does not lead to ambiguities, e.g. $p_I\equiv I$ or $k_j=j$. Throughout
this paper lower case Latin labels are used for light-like vectors and
upper case Latin labels for massive vectors.}
\begin{subequations}
\begin{align}
\Pi_+u(k_i)=\Pi_+v(k_i)&=\kea{i},&\Pi_-u(k_i)=\Pi_-v(k_i)&=\keb{i},\\
\bar{u}(k_i)\Pi_-=\bar{v}(k_i)\Pi_-&=\brb{i},&%
\bar{u}(k_i)\Pi_+=\bar{v}(k_i)\Pi_+&=\bra{i}.
\end{align}
\end{subequations}
Also in the massive case we use a similar bracket notation to distinguish the
different solutions of the Dirac equation.
We notice that using a given lightlike vector $q$ every massive
vector $p_I$ can be decomposed into a sum of two lightlike vectors
as in the following equation
\begin{equation}\label{eq:lcvectors}
p_I^\mu=k_i^\mu+\frac{(p_I)^2}{2p_I\cdot q}q^\mu
\end{equation}
which defines the lightlike vector $k_i$. 
The solutions of the Dirac equations
$(\pslash[I]\pm m_I)\kea{I^\pm}=0$ and $(\pslash[I]\pm m_I)\keb{I^\pm}=0$
can be expressed in terms of the massless spinors of $q$ and $k_i$,
\begin{subequations}\label{eq:lcspinors}
\begin{align}
\kea{I^\pm}&=\kea{i}\pm\frac{m_I}{\Spbb{iq}}\keb{q},\quad
&\keb{I^\pm}=\keb{i}\pm\frac{m_I}{\Spaa{iq}}\kea{q}\text{,}\\
\bra{I^\pm}&=\bra{i}\pm\frac{m_I}{\Spbb{qi}}\brb{q},\quad
&\brb{I^\pm}=\brb{i}\pm\frac{m_I}{\Spaa{qi}}\bra{q}.
\end{align}
\end{subequations}
It should be noted that in spite of the similarity in notation the
massive spinors are not constructed as eigenstates of the helicity
projectors~$\Pi_\pm$.

\subsection{Dimension Splitting}
In this section we want to derive the formul\ae{} required for separating
the dependence on $\tilde{g}$ from the spinor chains and spinor traces.
The starting point for such a separation is the following equation,
a proof of which can be found in~\cite{Reiter:2009kb}:
\begin{equation}\label{eq:SplitTracesD}
\tr{\One}\tr{\hat{\gamma}^{\mu_1}\cdots\hat{\gamma}^{\mu_p}
\tilde{\gamma}^{\nu_1}\cdots\tilde{\gamma}^{\nu_q}}=
\tr{\hat{\gamma}^{\mu_1}\cdots\hat{\gamma}^{\mu_p}}
\tr{\tilde{\gamma}^{\nu_1}\cdots\tilde{\gamma}^{\nu_q}}
\end{equation}
A chain of Dirac matrices can always be sorted by splitting
$\gamma^\mu=\hat{\gamma}^\mu+\tilde{\gamma}^\mu$ and shuffling
$\hat{\gamma}$ and $\tilde{\gamma}$ to opposite ends of the chain
using Equation~\eqref{eq:tHoofAlgebra}. The presence of $\gamma_5$
does not change the above results as in our scheme it can be
written as $\gamma_5=i\epsilon_{\mu\nu\rho\sigma}%
\hat{\gamma}^\mu\hat{\gamma}^\nu\hat{\gamma}^\rho\hat{\gamma}^\sigma/4!$.

Equation~\eqref{eq:SplitTracesD} extends to the case of spinor chains
by the observation that for any pair of massless spinors delimiting a
product of Dirac matrices one can introduce light-like auxiliary vectors
$p$ and $q$ in order to turn the spinor chain into a trace.
\begin{equation}\label{eq:glueinsertions}
\begin{array}{rlrl}
\Spab{i\vert\cdots\vert j}&=%
\frac{\displaystyle\tr{\Pi_-\kslash[i]\cdots\kslash[j]\pslash}}{%
\displaystyle\Spab{j\vert\pslash\vert i}},%
&\Spaa{i\vert\cdots\vert j}&=%
\frac{\displaystyle\tr{\Pi_-\kslash[i]\cdots\kslash[j]\pslash\qslash}}{%
\displaystyle\Spbb{j\vert\pslash\qslash\vert i}}%
\\\hphantom{-}&&&\\
\Spba{i\vert\cdots\vert j}&=%
\frac{\displaystyle\tr{\Pi_+\kslash[i]\cdots\kslash[j]\pslash}}{%
\displaystyle\Spba{j\vert\pslash\vert i}},%
&\Spbb{i\vert\cdots\vert j}&=%
\frac{\displaystyle\tr{\Pi_+\kslash[i]\cdots\kslash[j]\pslash\qslash}}{%
\displaystyle\Spaa{j\vert\pslash\qslash\vert i}}
\end{array}
\end{equation}
In all four cases, the insertion can be undone after
Equation~\eqref{eq:SplitTracesD} has been applied and hence no spurious
denominators need to be inserted in an actual calculation.

The trace of $(n-4)$ dimensional Dirac matrices $\tilde{\gamma}$ can be
evaluated according to the recursion relation
\begin{equation}
\tr{\tilde{\gamma}^{\nu_1}\cdots\tilde{\gamma}^{\nu_q}}=
\sum_i=2^q(-1)^i\tilde{g}^{\nu_1\nu_i}
\tr{\tilde{\gamma}^{\nu_2}\cdots\tilde{\gamma}^{\nu_{i-1}}%
\tilde{\gamma}^{\nu_{i+1}}\cdots\tilde{\gamma}^{\nu_q}}.
\end{equation}
Since only even numbers of $\tilde{\gamma}$ lead to non-vanishing
contributions, a chain with two Dirac matrices can be expanded
according to the above rules into
\begin{equation}\label{eq:tHooft-example01}
\Spaa{i\vert\gamma^\mu\gamma^\nu\vert j}=
\Spaa{i\vert\hat{\gamma}^\mu\hat{\gamma}^\nu\vert j}+
\Spaa{i\vert\tilde{\gamma}^\mu\tilde{\gamma}^\nu\vert j}=
\Spaa{i\vert\hat{\gamma}^\mu\hat{\gamma}^\nu\vert j}+
\tilde{g}^{\mu\nu}\Spaa{ij}
\end{equation}
As a more complicated example, we also give an explicit expression
for a spinor chain with four matrices:
\begin{multline}\label{eq:tHooft-example02}
\Spbb{i\vert\gamma^\mu\gamma^\nu\gamma^\rho\gamma^\sigma\vert j}=
\Spbb{i\vert\hat{\gamma}^\mu\hat{\gamma}^\nu%
\hat{\gamma}^\rho\hat{\gamma}^\sigma\vert j}
+\tilde{g}^{\rho\sigma}\Spbb{i\vert\hat{\gamma}^\mu\hat{\gamma}^\nu\vert j}
-\tilde{g}^{\nu\sigma}\Spbb{i\vert\hat{\gamma}^\mu\hat{\gamma}^\rho\vert j}\\
+\tilde{g}^{\nu\rho}\Spbb{i\vert\hat{\gamma}^\mu\hat{\gamma}^\sigma\vert j}
+\tilde{g}^{\mu\sigma}\Spbb{i\vert\hat{\gamma}^\nu\hat{\gamma}^\rho\vert j}
-\tilde{g}^{\mu\rho}\Spbb{i\vert\hat{\gamma}^\nu\hat{\gamma}^\sigma\vert j}
+\tilde{g}^{\mu\nu}\Spbb{i\vert\hat{\gamma}^\rho\hat{\gamma}^\sigma\vert j}\\
+\Spbb{ij}\left(
 \tilde{g}^{\mu\nu}\tilde{g}^{\rho\sigma}
-\tilde{g}^{\mu\rho}\tilde{g}^{\nu\sigma}
+\tilde{g}^{\mu\sigma}\tilde{g}^{\nu\rho}\right)\text{.}
\end{multline}

\subsection{Chisholm Identities}\label{ssec:chisholm}
With the same argument as in the previous section, we can extend the
validity of the Chisholm identity
\begin{equation}
\tr{\hat{\gamma}^{\mu_1}\cdots\hat{\gamma}^{\mu_{2m-1}}\hat{\gamma}^\nu}%
\hat{\gamma}_\nu=2\left(
\hat{\gamma}^{\mu_1}\cdots\hat{\gamma}^{\mu_{2m-1}}+
\hat{\gamma}^{\mu_{2m-1}}\cdots\hat{\gamma}^{\mu_1}
\right)
\end{equation}
to the case of spinor chains by substituting
Equation~\eqref{eq:glueinsertions}. If we denote products of
Dirac matrices as $\Gamma=\hat{\gamma}^{\mu_1}\cdots\hat{\gamma}^{\mu_p}$
and $\Gamma^\prime=\hat{\gamma}^{\nu_1}\cdots\hat{\gamma}^{\nu_q}$ and
their reversed strings by
$\overleftarrow{\Gamma}=\hat{\gamma}^{\mu_p}\cdots\hat{\gamma}^{\mu_1}$,
the Chisholm identities for spinor chains read
\begin{subequations}\label{eq:chisholmsp}
\begin{align}
\label{eq:chisholmsp-aa}
\Spaa{i\vert\Gamma\hat{\gamma}^\mu\Gamma^\prime\vert j}\cdot\hat{\gamma}_\mu
&=2\Gamma^\prime\kea{j}\bra{i}\Gamma
 -2\overleftarrow{\Gamma}\kea{i}\bra{j}\overleftarrow{\Gamma^\prime}\\
\label{eq:chisholmsp-bb}
\Spbb{i\vert\Gamma\hat{\gamma}^\mu\Gamma^\prime\vert j}\cdot\hat{\gamma}_\mu
&=2\Gamma^\prime\keb{j}\brb{i}\Gamma
 -2\overleftarrow{\Gamma}\keb{i}\brb{j}\overleftarrow{\Gamma^\prime}\\
\label{eq:chisholmsp-ab}
\Spab{i\vert\Gamma\hat{\gamma}^\mu\Gamma^\prime\vert j}\cdot\hat{\gamma}_\mu
&=2\Gamma^\prime\keb{j}\bra{i}\Gamma
 +2\overleftarrow{\Gamma}\kea{i}\brb{j}\overleftarrow{\Gamma^\prime}\\
\label{eq:chisholmsp-ba}
\Spba{i\vert\Gamma\hat{\gamma}^\mu\Gamma^\prime\vert j}\cdot\hat{\gamma}_\mu
&=2\Gamma^\prime\kea{j}\brb{i}\Gamma
 +2\overleftarrow{\Gamma}\keb{i}\bra{j}\overleftarrow{\Gamma^\prime}
\end{align}
\end{subequations}
The above identities are valid only if the number of Dirac matrices
in $\Gamma$ and $\Gamma^\prime$ matches the helicities of the spinors
$i$ and $j$, i.e. in equations \eqref{eq:chisholmsp-aa} and \eqref{eq:chisholmsp-bb}
the length of $\Gamma\hat\gamma^\mu\Gamma^\prime$ must be even\footnote{
$\gamma_5$ counts as an even number of matrices} and in
\eqref{eq:chisholmsp-ab} and \eqref{eq:chisholmsp-ba} it must be odd.

The repeated application of Equation~\eqref{eq:chisholmsp}
together with the identities
\begin{equation}
\hat{\gamma}^\mu\Gamma\hat{\gamma}_\mu=-2\overleftarrow{\Gamma}%
\quad\text{and}\quad%
\hat{\gamma}^\mu\Gamma\hat{\gamma}^\nu\hat{\gamma}_\mu=2\left(%
\hat{\gamma}^\nu\Gamma+\overleftarrow{\Gamma}\hat{\gamma}^\nu\right)
\end{equation}
for an odd number of matrices in $\Gamma$,
ensures that, starting from an expression
with no uncontracted Lorentz indices, all spinor chains in four dimensions
and traces can be expressed in terms of spinor products of the form
$\Spaa{ij}$ and~$\Spbb{ij}$.

\subsection{Majorana Spinors}\label{ssec:majo-th}
In supersymmetric extensions of the standard model
we need to deal with interactions of Majorana fermions.
A Majorana fermion is its own anti-particle,
i.e. it is invariant under charge conjugation:
\begin{equation}
 \widetilde{\psi}_{M} = C \bar{\psi}_{M}^{T} = \psi_{M}.
\end{equation}
The charge-conjugation matrix has the properties:
\begin{equation}\label{eq:Cproperties}
 C^{\dagger} = C^{-1}, \quad C^{T} = - C, \quad C \Gamma^{T}_{i} C^{-1} = \eta_{i} \Gamma_{i} 
\end{equation}
with
\begin{align}\label{eq:etaproperties}
\eta_{i} = 
\left\{ 
\begin{array}{l}
 +1 \mbox{ for } \Gamma_{i} = 1, \gamma_{5}, \gamma_{\mu}\gamma_{5} \\
 -1 \mbox{ for } \Gamma_{i} = \gamma_{\mu}.
\end{array} 
\right.
\end{align}

We aim to write down a consistent set of Feynman rules to deal with Majorana
fermions. The problem is that vertices involving Majorana fermions
violate fermion number flow; the fermion flow in the Feynman diagram is
ill-defined.
A consistent way of dealing with this was proposed
in~\cite{Haber:1984rc}.
There are two drawbacks with this approach; charge conjugation matrices are
explicitly introduced into the Feynman rules and the relative sign
of the Feynman graphs needs to be determined from the original Wick contractions.

We follow the approach in~\cite{Denner:1992vza}
which has been implemented in~\cite{Hahn:2000kx}. 
Each vertex containing Dirac fermions has two expressions,
one in which the fermion flow follows the fermion
number flow, and the other ``flipped'' vertex,
where the fermion flows in the
opposite direction to the fermion number flow.
In our implementation we impose a fermion flow on the vertex 
through the procedure \verb|RemoveNCContainer| and
``flip'' the vertices and spinors where neccesary.

We have implemented the following flipping rules, using~\eqref{eq:etaproperties}:
\begin{equation}\label{eq:fliprules}
\begin{array}{cc}
(\gamma^{\mu})^{'} &= -\gamma^{\mu}  \\
(\Pi_{\pm})^{'} &= \Pi_{\pm}
\end{array}
\end{equation}
and for the spinors 
\begin{equation}~\label{eq:spinorCtransform}
\begin{array}{cc}
\kea{p}^{'} &= \bra{p} \\
\keb{p}^{'} &= \brb{p}.
\end{array}
\end{equation}

This leads to the following rule for the
fermionic propagator:
\begin{equation}
S^{'}(p) = \frac{1}{-\pslash - m} = S(-p).
\end{equation} 

Another appealing feature of this method is that
the relative sign between graphs can be determined directly from the expressions
as opposed to reverting back to the original Wick contractions.
This brings it in line with the usual formulation for Dirac fermions.
To compute a consistent relative sign of Feynman graphs each must be multiplied by
$(-1)^{P+L}$, with, 
\begin{itemize}\label{RSIFlist}
 \item{P: the parity of the permutation of external spinors with respect to some reference order,}
\item{L: the number of closed fermion loops}.
\end{itemize}
This sign is determined after the flipping rules are applied.

\section{Program Description}\label{sec:interface}

\subsection{Installation}
The library has been written using the literate programming tool
\texttt{nuweb}~\cite{nuweb}. It can be obtained by downloading the file
\texttt{spinney.tgz} from the URL
\verb!http://www.nikhef.nl/~thomasr/filetransfer.php=spinney.tgz!\ .
The tarball contains the following files:
\begin{description}
\item[\texttt{spinney.hh}] the \FORM{} file
\item[\texttt{spinney.pdf}] annotated source code
\item[\texttt{spinney.nw}] the \texttt{nuweb} sources
\item[\texttt{spinney\_test.frm}] unit tests
\end{description}
After unpacking the library the should run the test program
\texttt{spinney\_test.frm} in order to ensure that the installed
\FORM{} version is recent enough. If one or more tests in the
program fail the most likely reason is a deprecated version of \FORM{}.

The file \texttt{spinney.nw} is only needed if the user wants to rebuild
any of the other files from scratch.

\subsection{Representation in \FORM}\label{ssec:representations}
The \texttt{spinney} library uses four different representations for
spinorial objects, each of which is considered the most convenient form at
a given point in the workflow.

For the description of the definitions and declarations we use the
\FORM{} keywords
(\textbf{CFunction}, \textbf{NFunction}, \textbf{CTensor}, \textbf{NTensor},
\textbf{Vector}, \textbf{Symbol}, \textbf{\#Procedure}) in the text to 
indicate its type. Where more than one definition is combined in one box
the arguments apply to all of the definitions.

\subsubsection{Non-Commuting Objects}
\label{ssec:noncommob}
\lstset{language={[spinney]form}}
The 't~Hooft algebra and the dimension splitting are applied to
\FORM{}'s non-commuting object. In order to keep the information about
the dimension in the function names rather than the indices we do not
use \FORM{}'s Dirac matrices here (\lstinline!g_!) but introduce our
own functions.

\medskip
\begin{NTensor}{Sm}{$\mu$}\index{functions!Sm|main}
	An $n$-dimensional Dirac matrix~$\gamma^\mu$.

	\begin{Parameters}
		\begin{Ventry}{$\mu$}
			\item[$\mu$] Lorentz index or vector
		\end{Ventry}
	\end{Parameters}
\end{NTensor}

\medskip
\begin{NTensor}{Sm4, SmEps}{$\mu$}
	\index{functions!Sm4|main}\index{functions!SmEps|main}
	The $4$-dimensional and $(n-4)$-dimensional (resp.) projections
	$\hat{\gamma}^\mu$ and $\tilde{\gamma}^\mu$
	of the Dirac matrix $\gamma^\mu$.
\end{NTensor}

\medskip
The non-commuting objects corresponding to the four-dimensional
matrix $\gamma_5$ and the derived projectors $\Pi_\pm=(\One\pm\gamma_5)/2$
are described below; they have been implemented as non-commuting functions
without parameters.

\medskip
\begin{NFunction}{Gamma5}{}
	\index{functions!Gamma5|main}
   The Dirac matrix $\gamma_5$.
\end{NFunction}

\medskip
\begin{NFunction}{ProjPlus, ProjMinus}{}
	\index{functions!ProjPlus|main}\index{functions!ProjPlus|main}
   The projectors $\Pi_R=\Pi_+=(\One+\gamma_5)/2$ and
	$\Pi_L=\Pi_-=(\One-\gamma_5)/2$ (resp.)
\end{NFunction}

\medskip
In the case of the spinors we use the same function names for both massive
and massless spinors. Massive spinors are distinguished by their second
argument~$\rho$, which is $\pm1$, indicating that the spinor lies in the
kernel of the operator $(\pslash-\rho m)$. For massless spinors the second
argument must be omitted.

\medskip
\begin{NFunction}{USpa, UbarSpb}{$p$ [, $\rho$]}
	\index{functions!USpa|main}\index{functions!UbarSpb|main}
   The massless spinor $\kea{p}$ and its conjugated $\brb{p}$,
	if $\rho$ is omitted;
	the massive spinor~$\kea{p^\rho}$ and its conjugated~$\brb{p^\rho}$
	otherwise.

	\begin{Parameters}
		\begin{Ventry}{$\rho$}
			\item[$p$] momentum of the spinor
			\item[$\rho$] (optional), a value $\pm1$ as described above
		\end{Ventry}
	\end{Parameters}
\end{NFunction}

\medskip
\begin{NFunction}{USpb, UbarSpa}{$p$ [, $\rho$]}
	\index{functions!USpb|main}\index{functions!UbarSpa|main}
   The massless spinor $\keb{p}$ and its conjugated~$\bra{p}$,
	if $\rho$ is omitted;
	the massive spinor~$\keb{p^\rho}$ and its conjugated $\bra{p^\rho}$
	otherwise.

	\begin{Parameters}
		\begin{Ventry}{$\rho$}
			\item[$p$] momentum of the spinor
			\item[$\rho$] (optional), a value $\pm1$ as described above
		\end{Ventry}
	\end{Parameters}
\end{NFunction}

\medskip
The following two functions mark the beginning and the end
of a Dirac trace.

\medskip
\begin{NFunction}{trL, trR}{}
	\index{functions!trL|main}\index{functions!trR|main}
   Indicates the begin (resp. end) of a Dirac trace.
\end{NFunction}

\medskip
For example, the expression $\tr{\Pi_+\gamma^\mu\pslash\gamma^\nu\qslash}$
would correspond to the following product in \FORM{}:
\begin{spform}
trL * ProjPlus * Sm(mu) * Sm(p) * Sm(nu) * Sm(q) * trR
\end{spform}

\subsubsection{Indexed Notation}
In this form the spinor indices of Dirac matrices and spinors are kept.
For example, the term $\Spab{k_1\vert\pslash\vert k_2}$ would be expressed as
\begin{displaymath}
(\bra{k_1})_{\alpha_1} (\pslash)_{\alpha_1\alpha_2} (\keb{k_2})_{\alpha_2}
\end{displaymath}
and in a \FORM{} program as
\begin{spform}
NCContainer(UbarSpa(k1), alpha1) *
NCContainer(USpb(k2), alpha2) *
NCContainer(Sm(p), alpha1, alpha2)
\end{spform}

This notation is particularly useful when importing expressions from a
diagram generator, which does not necessarily put the factors in the
correct order. All elements of the spinor line are wrapped inside
the function \texttt{NCContainer} which is defined as follows.

\medskip
\begin{CFunction}{NCContainer}{$o$, $i_1$ [, $i_2$]}
	\index{functions!NCContainer|main}
	Representation of a spinor or a Dirac matrix with explicit
	spinor indices.

	\begin{Parameters}
		\begin{Ventry}{$i_1$}
			\item[$o$] a non-commuting object, such as \texttt{Sm},
				\texttt{Gamma5} or \texttt{UbarSpa}; a complete list
				is given in Section~\ref{ssec:noncommob}.
			\item[$i_1$] for spinors: the only spinor index;
			   for Dirac matrices: the first spinor index
			\item[$i_2$] for spinors: not present;
			   for Dirac matrices: the second spinor index
		\end{Ventry}
	\end{Parameters}
\end{CFunction}

\subsubsection{Collected Form}
The collected form is generated from non-commutative objects
by the procedure \texttt{SpCollect} or from the open form by
the procedure \texttt{SpClose}. Spinor strings are represented by
the following four functions

\medskip
\begin{CFunction}{Spaa, Spab, Spba, Spbb}{$k_1$, \dots, $k_2$}
	\index{functions!Spaa|main}\index{functions!Spab|main}
	\index{functions!Spba|main}\index{functions!Spbb|main}
   A string delimited by two spinors. The four forms correspond
	to $\Spaa{k_1\vert\ldots\vert k_2}$, $\Spab{k_1\vert\ldots\vert k_2}$,
   $\Spba{k_1\vert\ldots\vert k_2}$ and $\Spbb{k_1\vert\ldots\vert k_2}$
	respectively.

	\begin{Parameters}
		\begin{Ventry}{$\ldots$}
			\item[$k_1$] a light-like, four-dimensional vector.
			\item[$\ldots$] a list of four-dimensional indices or vectors.
			\item[$k_2$] a light-like, four-dimensional vector.
		\end{Ventry}
	\end{Parameters}
\end{CFunction}

\subsubsection{Open Form}
The open form can be generated from the collected form by
calling the procedure \texttt{SpOpen} after all Lorentz indices
have been contracted using~\texttt{SpContract}. The reverse operation
of \texttt{SpOpen} is \texttt{SpClose}, which transforms the open
form back to collected form.

\medskip
\begin{CFunction}{Spa2, Spb2}{$k_1$, $k_2$}
	\index{functions!Spa2|main}\index{functions!Spb2|main}
   The spinor products $\Spaa{k_1k_2}$ and $\Spbb{k_1k_2}$ respectively.
	These functions are defined as anti-symmetric in their
	arguments.

	\begin{Parameters}
		\begin{Ventry}{$k_2$}
			\item[$k_1$] a light-like, four-dimensional vector.
			\item[$k_2$] a light-like, four-dimensional vector.
		\end{Ventry}
	\end{Parameters}
\end{CFunction}

\subsubsection{Metric Tensors}
The dimension splitting of $g^{\mu\nu}$ into the sum
$\hat{g}^{\mu\nu}+\tilde{g}^{\mu\nu}$ as described in
Equation~\eqref{eq:dimension-splitting} is reflected in the
definition of the metric tensor and its projections onto subspaces.
As in the case of the Dirac matrices we find it more convenient to
define new functions for the metric tensor rather than using the
tensor \texttt{d\_}, which is predefined in Form.

\medskip
\begin{CTensor}{d}{$\mu$, $\nu$}
	\index{functions!d|main}
   The $n$-dimensional metric tensor $g^{\mu\nu}$.
	This function is defined symmetric in its arguments.
\end{CTensor}

\medskip
\begin{CTensor}{d4, dEps}{$\mu$, $\nu$}
	\index{functions!d4|main}\index{functions!dEps|main}
   \texttt{d4} is the $4$-dimensional projection $\hat{g}^{\mu\nu}$
	of the metric tensor;
	\texttt{dEps} is the orthogonal projection $\tilde{g}^{\mu\nu}$ into the
	$(n-4)$ dimensional subspace.

	These function are defined symmetric in their arguments.
\end{CTensor}

\subsubsection{Other Symbols and Objects}
The function \texttt{SpDenominator} has been introduced in order to represent
the reciprocal value of its argument. Therefore, the replacement
$\mathtt{SpDenominator}(x\mathtt{?})\rightarrow 1/x$ is always safe.
However, keeping certain denominators inside function arguments makes
some substitutions easier.

\medskip
\begin{CFunction}{SpDenominator}{$x$}
	\index{functions!SpDenominator|main}
   Represents the reciprocal value of the argument, i.e.~$1/x$.
\end{CFunction}
\medskip

The function \texttt{SpERRORTOKEN} has been introduced to indicate
inconsitencies detected by the procedure \texttt{SpCheck} which
signals if the length of a spinor string trivially nullifies an
expression. Although these null-expressions are not errors, in certain
circumstances they can indicate errors in a program.

\medskip
\begin{CFunction}{SpERRORTOKEN}{}
	\index{functions!SpERRORTOKEN|main}
   Indicates that the procedure \texttt{SpCheck} found an error
	in the term where \texttt{SpERRORTOKEN} appears.

\end{CFunction}

\subsection{Reserved Symbols}
The library \texttt{spinney} defines more objects for internal use.
All implemented algorithms require that none of these objects are
present in any active expression at the invocation of the procedure and
ensure that these objects are not present in any active expression after
the procedure returns. These objects are declared exactly as written below.
\begin{spform}
CFunctions fDUMMY1, ..., fDUMMY4;
Symbols sDUMMY1, ..., sDUMMY4;
Indices iDUMMY1, ..., iDUMMY4;
Vectors vDUMMY1, ..., vDUMMY4;
NFunctions nDUMMY1, SpFlip;
\end{spform}
We also define the three sets \texttt{SpORIGSet}, \texttt{SpIMAGSet}
and \texttt{SpObject} which are used inside \texttt{RemoveNCContainer}.

\subsection{Implemented Algorithms}

\subsubsection{Light-Cone Decomposition}
This algorithm uses Equations~\eqref{eq:lcvectors}
and~\eqref{eq:lcspinors} in order to express massive
vectors and spinors in terms of massless ones.

\medskip
\begin{Procedure}{LightConeDecomposition}{$p_I$, $p_i$, $q$, $m_I$}
	In every active expression all appearances of $p_I$ and all
	spinors $\kea{I^\pm}$, $\keb{I^\pm}$, $\bra{I^\pm}$ and
	$\brb{I^\pm}$ are replaced by a pair of light-like vectors
	(resp. spinors).

	The parameters are defined according to Equation~\eqref{eq:lcvectors}:

	\begin{Parameters}
		\begin{Ventry}{$m_I$}
			\item[$p_I$] a massive vector that fulfills $(p_I)^2=m_I^2$
			\item[$p_i$] a light-like vector
			\item[$q$] a light-like vector (reference momentum)
			\item[$m_I$] the mass of $p_I$
		\end{Ventry}
	\end{Parameters}

	This procedure works on non-commuting objects. If massive spinors
	are present this procedure must be called before \texttt{SpCollect}.
\end{Procedure}

\medskip
This procedure needs to be called before \texttt{SpCollect} as
in the closed form only massless spinors are allowed.

As an example, we write down the numerator of the color stripped
tree level diagram of
$u\bar{u}\rightarrow t\bar{t}$ in QCD for two different helicities.

\begin{spform}
Local d1 = UbarSpa(k1) * Sm4(mu) * USpb(k2) *
           UbarSpa(p3, +1) * Sm4(mu) * USpa(p4, -1);
Local d2 = UbarSpa(k1) * Sm4(mu) * USpb(k2) *
           UbarSpb(p3, +1) * Sm4(mu) * USpa(p4, -1);

#call LightConeDecomposition(p3, k3, k1, mT)
#call LightConeDecomposition(p4, k4, k2, mT)
#call SpCollect
#call SpContract

Print;
.end
\end{spform}
We have also added the command \texttt{SpContract} in this example
as it simplifies the output considerably, which is given below.
\begin{form}
   d1 = 2*Spaa(k1,k4)*Spbb(k1,k2)*
        SpDenominator(Spb2(k1,k3))*mT;
   d2 = 2*Spaa(k1,k4)*Spbb(k3,k2);
  0.00 sec out of 0.03 sec
\end{form}
The above calculation can therefore be summarized as:
\begin{align*}
d_1&=\Spab{k_1\vert\hat{\gamma}^\mu\vert k_2}%
     \Spaa{p_3^+\vert\hat{\gamma}_\mu\vert p_4^-}=
	  2\cdot\frac{\Spaa{k_1k_4}\Spbb{k_1k_2}}{\Spbb{k_1k_3}}\cdot m_T\\
d_2&=\Spab{k_1\vert\hat{\gamma}^\mu\vert k_2}%
     \Spab{p_3^+\vert\hat{\gamma}_\mu\vert p_4^-}=
	  2\cdot\Spaa{k_1k_4}\Spbb{k_3k_2}
\end{align*}

\subsubsection{$n$-Dimensional 't~Hooft Algebra}

\begin{Procedure}{tHooftAlgebra}{}
   Carries out the $(n-4)$-dimensional part of the algebra such
   that only four dimensional Dirac matrices $\hat{\gamma}^\mu$ are
   left; the dependence on $(n-4)$ is entirely expressed in terms
	of~$\tilde{g}^{\mu\nu}$.
	
	This procedure works on non-commuting objects. If $n$-dimensional
	Dirac matrices, $\Pi_\pm$ or $\gamma_5$ are present this procedure
	must be called before \texttt{SpCollect}.
\end{Procedure}

\medskip
As an example, we give a short program that reproduces
Equations~\eqref{eq:tHooft-example01} and~\eqref{eq:tHooft-example02}.
\begin{spform}
Local example1 =
   UbarSpa(k1) * Sm(mu) * Sm(nu) * USpa(k2);
Local example2 =
   UbarSpa(k1) * Sm(mu) * Sm(nu) *
      Sm(rho) * Sm(sigma) * USpa(k2);
#call tHooftAlgebra()
#call SpCollect
Print +s;
\end{spform}

\subsubsection{Change of Representation}
\label{sec:ChangeRep}
As described in Section~\ref{ssec:representations},
different representations for the
expressions are used at different points in a \FORM{}
program using~\spinney{}. Here, we give an overview of
the functions changing between the representations.
While for most representations only one-directional translation is
provided, the user can change forth and back between collected form
and open form, which is indicated in the diagram below
\begin{multline}
\mathtt{NCContainer(UbarSpa(p),i)*NCContainer(USpa(q),i)}
\stackrel{\mathtt{RemoveNCContainer}}{\longrightarrow}\\
\mathtt{UbarSpa(p)*USpa(b)}
\stackrel{\mathtt{SpCollect}}{\longrightarrow}
\mathtt{Spaa(p,q)}
{\mathop{\longleftrightarrow}\limits_{\texttt{SpClose}}^{\texttt{SpOpen}}}
\mathtt{Spa2(p,q)}
\end{multline}

The first in this collection of procedures is \texttt{RemoveNCContainer}.
Its effect amounts to stripping off the function \texttt{NCContainer}
and ordering the Dirac matrices and spinors according to the
order of the spinor indices. Where necessary the flipping rules for
Majorana fermions are applied (see Section~\ref{ssec:majo-th}).

\medskip
\begin{Procedure}{RemoveNCContainer}{}
Translates the indexed notation into non-commuting objects.
\end{Procedure}
\medskip

At the level of non-commuting objects the dimension splitting and the
't Hooft algebra take place, which has been implemented in the routine
\texttt{tHooftAlgebra}.

After these steps, typically, one wants to proceed in collected form.
This form is better suited for contractions across different spinor lines,
as in $\Spab{p_1\vert\hat{\gamma}^\mu\vert p_2}%
\Spab{p_3\vert\hat{\gamma}_\mu\vert p_4}$, which can be detected very
easily using argument lists but is very difficult to be carried out
using products of non-commuting objects.

Traces of the form
$\mathtt{trL}\cdots\mathtt{trR}$ can be converted to collected
form using the procedure~\texttt{SpTrace4}, all remaining products
of non-commuting objects can be transformed by the
procedure~\texttt{SpCollect}.

\medskip
\begin{Procedure}{SpTrace4}{$k_1$, $k_2$, \dots}
Operates on non-commutative objects in products representing
a trace (\texttt{trL * \dots * trR}).

	\begin{Parameters}
		\begin{Ventry}{$k_1$, $k_2$}
			\item[$k_1$, $k_2$, \dots] an optional list of vectors.
			If specified, these vectors are assumed lightlike and traces
			are opened at those positions. If no such positions are found
			the trace is evaluated using \FORM{}'s \texttt{Trace4} command.
		\end{Ventry}
	\end{Parameters}

\end{Procedure}
\medskip

The following example demonstrates the difference in treatment between
light-like vectors and those which have not been specified as light-like.
\begin{spform}
Local test1 = trL*Sm4(k1)*Sm4(k2)*Sm4(k3)*Sm4(k4)*trR;
Local test2 = trL*Sm4(k1)*Sm4(k2)*Sm4(q)*Sm4(k4)*trR;
#call SpTrace4(q)
Print;
.end
* test1 = 4*k1.k2*k3.k4-4*k1.k3*k2.k4+4*k1.k4*k2.k3;
* test2 = Spab(q,k4,k1,k2,q)+Spba(q,k4,k1,k2,q);
\end{spform}
Very often specifying the list of light-like vectors leads to
more compact expressions, especially if the number of terms
generated by taking the trace is very big.

\medskip
\begin{Procedure}{SpCollect}{}
Translates non-commuting objects into collected form.

This procedure requires that the $n$-dimensional algebra has already
been carried out by a call to \texttt{tHooftAlgebra} and that all
massive spinors have been replaced using the routine
\texttt{LightConeDecomposition}.

As a result the all spinorial objects are expressed in terms of
the function \texttt{Spaa}, \texttt{Spab}, \texttt{Spba} and \texttt{Spab}.
\end{Procedure}
\medskip

Before one can go from collected form to open form one should eliminate
all Lorentz contractions by the use of \texttt{SpContract} and
\texttt{SpContractLeviCivita}. The latter one is only needed if 
the Levi-Civita symbol $\epsilon^{\mu\nu\rho\sigma}$ has been introduced
by taking a trace involving $\gamma_5$.

The conversion from collected to open form is performed by
the procedure \texttt{SpOpen}. The reverse operation, converting from
open form back to collected form is done by the procedure \texttt{SpClose}.

\medskip
\begin{Procedure}{SpOpen}{$k_1$, $k_2$, \dots}
Translates an expression in collected form into open form.

In open form, only the functions \texttt{Spa2} and \texttt{Spb2}
are used to express spinor products.

This procedure requires that all Lorentz indices inside spinor
lines have been removed using the routine \texttt{SpContract}.

	\begin{Parameters}
		\begin{Ventry}{$k_1$, $k_2$}
			\item[$k_1$, $k_2$, \dots] an optional list of vectors. If specified,
			the spinor lines are only opened at positions indicated
			by the given vectors. In this case a complete translation
			into \texttt{Spa2} and \texttt{Spb2} functions might not
			
		\end{Ventry}
	\end{Parameters}
\end{Procedure}

The parameter list of \texttt{SpClose} allows to close spinor strings
at the positions of certain vectors, indicated by the arguments.
In both cases, \texttt{SpOpen} and \texttt{SpClose}, the parameter
lists facilitate the use of this library for unitarity based methods,
as will be shown in the examples of Section~\ref{ssec:ex:unitarity}.

\medskip
\begin{Procedure}{SpClose}{$k_1$, $k_2$, \dots}
The inverse operation of \texttt{SpOpen}. Replaces the functions
\texttt{Spa2} and \texttt{Spb2} into the functions \texttt{Spaa},
\texttt{Spab} etc.

	\begin{Parameters}
		\begin{Ventry}{$k_1$, $k_2$}
			\item[$k_1$, $k_2$, \dots] an optional list of vectors. If specified,
			only those positions are closed which are indicated by
			the given vectors.
		\end{Ventry}
	\end{Parameters}

\end{Procedure}
\medskip

In some cases this operation requires that a spinor string is broken up
at another position. Consider the following example:
\begin{spform}
Local expression = Spab(q, k1, k2, k1, q);

#call SpOpen()
#call SpClose(q)
Print;
.end
* test = - Spab(k1,q,k1)*Spa2(k1,k2)*Spb2(k1,k2);
\end{spform}
However, it is not guaranteed that the operation always succeeds:
if we had written \lstinline!#call SpClose(q,k1)! instead of
\lstinline!#call SpClose(q)! the following result would have
come out, where the vector \texttt{q} still appears in the spinors.
\begin{spform}
* test = Spab(k2,k1,k2)*Spab(q,k1,q);
\end{spform}

\subsubsection{Contraction of Lorentz Indices}
This Section describes three routines which reduce the number of
explicit appearances of contracted Lorentz indices. The first type
of contractions, those involving only spinor lines, as in
$\Spab{p_1\vert\cdots\hat{\gamma}^\mu\cdots\vert p_2}
\Spab{p_3\vert\cdots\hat{\gamma}_\mu\cdots\vert p_4}$
or in
$\Spab{p_1\vert\cdots\hat{\gamma}^\mu\cdots\hat{\gamma}_\mu\cdots\vert p_2}$,
are treated by the procedure \texttt{SpContract}.

\medskip
\begin{Procedure}{SpContract}{}
Applies the equations of Section~\ref{ssec:chisholm} in order to
eliminate all Lorentz indices inside spinor lines. This procedure
works on expressions in collected form.
\end{Procedure}
\medskip

Contractions involving metric tensors are simplified by
the procedure \texttt{SpContractMetrics}{}. This procedure is
implemented such that it can deal with expressions in collected form
but also with non-commuting objects.

\medskip
\begin{Procedure}{SpContractMetrics}{}
Removes spurious appearances of the metric tensors $g^{\mu\nu}$,
$\hat{g}^{\mu\nu}$ and $\tilde{g}^{\mu\nu}$. Whereas
$\hat{g}^\mu_\mu=4$ is substituted immediately, all other
instances of the dimensions $\tilde{g}^\mu_\mu$ and
$g^\mu_\mu$ are not replaced.
\end{Procedure}
\medskip

The last one in this category of procedures is
\texttt{SpContractLeviCivita}. This routine uses the fact
that the Levi-Civita tensor $\epsilon^{\mu\nu\rho\sigma}$
can be written as a trace:
\begin{equation}
\epsilon^{\mu\nu\rho\sigma}=
-\frac{i}{4}\tr{\gamma_5
\hat{\gamma}^\mu\hat{\gamma}^\nu\hat{\gamma}^\rho\hat{\gamma}^\sigma}\text{.}
\end{equation}
The procedure considers two cases. If the Levi-Civita tensor is contracted
with a Dirac matrix inside a spinor string it uses the
Chisholm identity to simplify
\begin{equation}\label{eq:levi:01}
\epsilon^{\mu\nu\rho\sigma}\hat{\gamma}_\sigma=
-\frac{i}2\left(\Pi_+-\Pi_-\right)\left[
\hat{\gamma}^\mu\hat{\gamma}^\nu\hat{\gamma}^\rho
-\hat{\gamma}^\mu\hat{\gamma}^\nu\hat{\gamma}^\rho\right]
\end{equation}
The second case is the contraction of the Levi-Civita tensor with
a lightlike momentum. Here one can rewrite the $\epsilon$-tensor
as
\begin{equation}\label{eq:levi:02}
p_\mu\epsilon^{\mu\nu\rho\sigma}=
-\frac{i}4\left(
\Spba{p\vert\hat{\gamma}^\nu\hat{\gamma}^\rho\hat{\gamma}^\sigma\vert p}
-\Spab{p\vert\hat{\gamma}^\nu\hat{\gamma}^\rho\hat{\gamma}^\sigma\vert p}
\right)
\end{equation}

Products of multiple $\epsilon$-tensors are reduced by applying
the determinant relation
\begin{equation}\label{eq:levi:03}
\epsilon^{\mu_1\mu_2\mu_3\mu_4}\epsilon^{\nu_1\nu_2\nu_3\nu_4}=
\det{(\hat{g}^{\mu_i\nu_j})_{i,j=1}^4}
\end{equation}

\medskip
\begin{Procedure}{SpContractLeviCivita}{$k_1$, $k_2$, \dots}
   Eliminates Levi-Civita tensors as far as possible applying
	Equations \eqref{eq:levi:01}, \eqref{eq:levi:02}
	and~\eqref{eq:levi:03}.

	\begin{Parameters}
		\begin{Ventry}{$k_1$, $k_2$}
			\item[$k_1$, $k_2$, \dots] an optional list of vectors.
			If specified, these vectors are assumed lightlike.
		\end{Ventry}
	\end{Parameters}

\end{Procedure}

\subsubsection{Schouten Identity}
In analogy to the package \SAM~\cite{Maitre:2007jq}
we have implemented three versions of the Schouten identity
\begin{equation}
\Spaa{ij}\Spaa{kl}=\Spaa{il}\Spaa{kj}+\Spaa{ik}\Spaa{jl}
\end{equation}
and its conjugated version.

\begin{Procedure}{Schouten}{$p_1$, $p_2$, $q_1$, $q_2$}
Substitutes according to the Schouten identity
\begin{displaymath}
\Spaa{p_1p_2}\Spaa{q_1q_2}\rightarrow
\Spaa{p_1q_2}\Spaa{q_1p_2}+\Spaa{p_1q_1}\Spaa{p_2q_2}.
\end{displaymath}
\end{Procedure}

\medskip
\begin{Procedure}{Schouten}{$p_1$, $p_2$, $q$}
Substitutes according to the Schouten identity
\begin{displaymath}
\forall q^\prime: \Spaa{p_1p_2}\Spaa{qq^\prime}\rightarrow
\Spaa{p_1q^\prime}\Spaa{qp_2}+\Spaa{p_1q}\Spaa{p_2q^\prime}.
\end{displaymath}
\end{Procedure}

\medskip
\begin{Procedure}{Schouten}{$q$}
Substitutes according to the Schouten identity
\begin{displaymath}
\forall p_1,p_2,p_3: \frac{\Spaa{qp_1}}{\Spaa{qp_2}\Spaa{qp_3}}\rightarrow
\frac{\Spaa{p_2p_1}}{\Spaa{qp_2}\Spaa{p_2p_3}}
-\frac{\Spaa{p_3p_1}}{\Spaa{qp_3}\Spaa{p_2p_3}}.
\end{displaymath}
\end{Procedure}

\subsubsection{Miscellanous Routines}
In this section we describe some routines which are of less
common use. The procedure \texttt{SpClear} nullifies all expressions
which should vanish due to spinor lines of the wrong length.
The procedure \texttt{SpClear} is called inside \texttt{SpContract}
because Equations~\eqref{eq:chisholmsp} hold only if the spinor chains
have correct length.

\medskip
\begin{Procedure}{SpClear}{}
   Eliminates all terms which contain a spinor line
	$\Spaa{p\vert\Gamma\vert q}$ or $\Spbb{p\vert\Gamma\vert q}$,
	where $\Gamma$ is a product of an odd number of Dirac matrices,
	or a spinor line
	$\Spab{p\vert\Gamma\vert q}$ or $\Spba{p\vert\Gamma\vert q}$,
	where $\Gamma$ is a product of an even number of Dirac matrices.

	This procedure acts on expressions in collected form.
\end{Procedure}
\medskip

The routine \texttt{SpCheck}
acts in a similar way as \texttt{SpClear} does.
Instead of removing the terms which should vanish trivially
it marks them with the function \texttt{SpERRORTOKEN}.

\medskip
\begin{Procedure}{SpCheck}{}
   Substitutes all terms which contain a spinor line
	$\Spaa{p\vert\Gamma\vert q}$ or $\Spbb{p\vert\Gamma\vert q}$,
	where $\Gamma$ is a product of an odd number of Dirac matrices,
	or a spinor line
	$\Spab{p\vert\Gamma\vert q}$ or $\Spba{p\vert\Gamma\vert q}$,
	where $\Gamma$ is a product of an even number of Dirac matrices
	inside the arguments of the function \texttt{SpERRORTOKEN}.

	This procedure acts on expressions in collected form.
\end{Procedure}

The last procedure of this section, \texttt{SpOrder}, shuffles
Dirac matrices into a given order using the 't~Hooft-Veltman
algebra. The order is specified by the argument list of
\texttt{SpOrder}. This routine is used in the test programs
to bring the results into a canonical form which is necessary
in order to check that all tests hold.

\medskip
\begin{Procedure}{SpOrder}{$k_1$,$k_2$,\dots}
   Sorts all products of Dirac matrices according to the
	ordering specified by the argument list. This routine
	acts on non-commuting objects.

	\begin{Parameters}
		\begin{Ventry}{$k_1$, $k_2$}
			\item[$k_1$, $k_2$, \dots] a list of vectors.
			If $k_i$ is to the left of $k_j$ in the argument list
			then the procedure shuffles $\kslash[i]$ to the left
			of $\kslash[j]$ in the expression.
		\end{Ventry}
	\end{Parameters}
\end{Procedure}

\subsection{Working with Majorana Spinors}

\subsubsection{On the Relative Sign of Feynman Graphs}

The flipping rules have been tested in Golem-2{.}0
where the diagrams are generated by Qgraf~\cite{Nogueira:1991ex}. 
The relative sign calculated by Qgraf is incorrect when dealing with Majorana fermions. 
Here, we present a method to calculate it. Firstly we
calculate $(-1)^{P}$ using the following code:

\begin{form}
Function NCOrder;
Id fDUMMY1?{UbarSpa,UbarSpb}(vDUMMY1?) =
      NCOrder(vDUMMY1)*fDUMMY1(vDUMMY1);
Id fDUMMY1?{USpa,USpb}(vDUMMY1?) =
      fDUMMY1(vDUMMY1)*NCOrder(vDUMMY1);
#call tHooftAlgebra
#call SpCollect
ChainIn NCOrder;
AntiSymmetrize NCOrder;
Id NCOrder(?all) = 1;
\end{form}

We multiply our diagram by a non-commuting function of the external momenta which encodes the order
of the spinors in the diagram.
The arguments of this function are then
brought into \FORM's natural ordering. 
The exchange of any two arguments results in a minus sign.

Secondly we must determine $(-1)^{L}$ with L being the number of closed fermion
loops. This is easily done by counting the number of appearances of \texttt{trL},
as long as one ensures that the only source of spinor traces in the first
place are closed fermion loops; the easiest way of multiplying the
amplitude by the correct sign could be implemented as \texttt{Id trL = -trL}.

\subsubsection{Fixing Fermion Chain Order}

The method we have described relies on being able
to fix the fermion chain order.
This is achieved in the procedure \texttt{NCContainer}.
In our code an incoming (outgoing) Majorana fermion 
will initially be treated as in incoming (outgoing) Dirac fermion (as opposed to an anti-fermion).
When we join an incoming Majorana spinor with an incoming Dirac fermion or an outgoing Dirac anti-fermion,
one spinor will need to be flipped.
The same applies when an outgoing Majorana fermion is joined to an incoming anti-fermion or an outgoing fermion.
As an example we have an expression:
\begin{form}
NCContainer(UbarSpa(k1)*Sm(i1)*ProjPlus*UbarSpa(k2))
\end{form}
which is transformed using \verb|RemoveNCContainer| to
\begin{form}
NCContainer(UbarSpa(k1), Sm(i1), ProjPlus,
				SpFlip(UbarSpa(k2)))
\end{form} and then
using \eqref{eq:spinorCtransform} we have the fermion chain:
\begin{form}
UbarSpa(k1)*Sm(i1)*ProjPlus*USpb(k2).
\end{form}

This is the default behaviour of \spinney{}; one can prevent \spinney{}
from applying the flipping rules by defining the preprocessor
variable \texttt{NOSPFLIP} before calling \texttt{RemoveNCContainer}.

\section{Examples and Applications}\label{sec:examples}
\subsection{Feynman Diagram Based Reduction of One-Loop Amplitudes}
In this section we show how the \FORM{} library \spinney{} can be used
for the reduction of one-loop diagrams, both with conventional reduction of
tensor integrals and with a reduction at the integrand level as described
in~\cite{Ossola:2006us}. In the first case we use the conventions
of~\cite{Binoth:2005ff}. However, the method is not restricted to to this
particular tensor decomposition.

\begin{figure}[tbph]
\begin{center}
\includegraphics{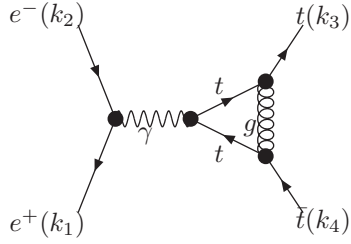}
\end{center}
\caption{Three-point diagram ($e^+e^-\rightarrow t\bar{t}$)
   discussed in the example.}
\label{fig:pyxovirt}
\end{figure}
In our setup, we use the diagram generator QGraf~\cite{Nogueira:1991ex}.
For the diagram in Figure~\ref{fig:pyxovirt} we obtain an output similar
to the one given below.\footnote{For the sake of simplicity we skip the
discussion of the color algebra.}
\begin{form}
Local diagram1 =
   inp(-1, iv1r1L1, k1, me) *
   inp(+1, iv1r2L1, k2, me) *
   out(+1, iv2r1L1, k3, mT) *
   out(-1, iv3r2L1, k4, mT) *
   vertex([field.ep], k1, iv1r1L1,
          [field.em], k2, iv1r2L1,
          [field.A], -k1-k2, iv1r3L2) *
   vertex([field.Ubar], -k3, iv2r1L1,
          [field.U], Q+k3, iv2r2L1,
          [field.g], -Q, iv2r3L2) *
   vertex([field.Ubar], -Q+k4, iv3r1L1,
          [field.U], -k4, iv3r2L1,
          [field.g], Q, iv3r3L2) *
   vertex([field.Ubar], -Q-k3, iv4r1L1,
          [field.U], Q-k4, iv4r2L1,
          [field.A], k1+k2, iv4r3L2) *
   prop(+2, -k1-k2, 0, iv4r3L2, iv1r3L2) *
   prop(+2, -Q, 0, iv3r3L2, iv2r3L2) *
   prop(+1, Q+k3, mT, iv4r1L1, iv2r2L1) *
   prop(+1, Q-k4, mT, iv3r1L1, iv4r2L1);
\end{form}

The first step is the substitution of the Feynman rules.
In the replacement of the wave functions the approximation $m_e=0$
is applied and helicities are assigned to the particles.
\begin{spform}
Id inp(-1,iv1?,k1,me)=NCContainer(UbarSpa(k1),iv1);
Id inp(+1,iv1?,k2,me)=NCContainer(USpb(k2),iv1);
Id out(+1,iv1?,k3,mT)=NCContainer(UbarSpb(k3,+1),iv1);
Id out(-1,iv1?,k4,mT)=NCContainer(USpa(k4,-1),iv1);
\end{spform}
As discussed in Section~\ref{ssec:noncommob} the letters `\texttt{a}'
and `\texttt{b}' distinguish the helicities, whereas the parameter $\pm1$
at the right-hand side differentiates between $u$- and $v$-spinors.

In a similar manner the propagators and vertices are substituted. Again,
focussing on the Lorentz algebra we omit color structure, coupling constants
and denominators here.
\begin{spform}
Id prop(1, k1?, mT?, iv1?, iv2?) =
   i_ * (NCContainer(Sm(k1), iv2, iv1)
      + mT * NCContainer(1, iv2, iv1));
Id prop(2, k1?, 0, iv1?, iv2?) = - i_ * d(iv1, iv2);
Id vertex([field.ep]?, k1?, iv1L1?,
          [field.em]?, k2?, iv2L1?,
          [field.A]?,  k3?, iv3L2?) =
   NCContainer(Sm(iv3L2), iv1L1, iv2L1);
\end{spform}
Now, a call to \texttt{RemoveNCContainer} brings the non-commuting
object in the right order and simplifies the representation of the
expression. We replace the massive spinors by projecting onto massless
vectors. Hereby we introduce new vectors $l_3$ and $l_4$ as defined
in Equations \eqref{eq:lcvectors} and~\eqref{eq:lcspinors}.
\begin{spform}
#call RemoveNCContainer
#call LightConeDecomposition(k3,l3,k1,mT)
#call LightConeDecomposition(k4,l4,k2,mT)
\end{spform}

At this point the manipulation of the diagram branches depending
on the output one wants to achieve. In a conventional reduction of
the tensor integrals the algorithm contiues after replacing the
integration momentum $Q$ by the corresponding expressions for the
tensor integrals. In the case of a reduction at the integrand level
we like to express $Q$ in terms of a four-dimensional projection
$Q_4$ and a scale $\mu^2$ such that $Q^2=(Q_4)^2-\mu^2$.

\subsubsection{Conventional Tensor Reduction}
The expression corresponding to the Feynman diagram in
Figure~\ref{fig:pyxovirt} has integrals of rank two at most.
Following the notation of~\cite{Binoth:2005ff} the momenta
in the loop are $r_1=-k_4$, $r_2=k_3$ and $r_3=0$. The corresponding
form factor representation of the occurring tensor integrals is
\begin{align}
I_3^n(S) &= A^{3,0}(S);\\
I_3^{n,\mu_1}(S) &= -k_4^{\mu_1} A^{3,1}_1(S) +k_3^{\mu_1} A^{3,1}_2(S);\\
I_3^{n,\mu_1\mu_2}(S) &=
      k_4^{\mu_1}k_4^{\mu_2} A^{3,2}_{11}(S)
    - (k_3^{\mu_1}k_4^{\mu_2}+k_4^{\mu_1}k_3^{\mu_2}) A^{3,2}_{12}(S)
\nonumber\\ &
    + k_4^{\mu_1}k_4^{\mu_2} A^{3,2}_{22}(S)
	 + g^{\mu_1\mu_2}B^{3,2}(S).
\end{align}

These formulae are easily implemented for the given example.
It should be noted that the calls to the function
\texttt{LightConeDecomposition} must be repeated
after this step since we reintroduce the momenta $k_3$ and $k_4$
here.
\begin{spform}
ToTensor, Functions, Q, QTens;
If(count(QTens,1)==0) Multiply A30;
Id QTens(iv1?) = -k4(iv1) * A31(1) + k3(iv1) * A31(2);
Id QTens(iv1?,iv2?) =
   + k3(iv1)*k3(iv2) * A32(1,1)
   - k3(iv1)*k4(iv2) * A32(1,2)
   - k4(iv1)*k3(iv2) * A32(1,2)
   + k4(iv1)*k4(iv2) * A32(2,2)
   + d(iv1,iv2) * B32();
\end{spform}

The remaining steps of the program carry out the 't~Hooft algebra
and simplify the expression by taking out all contractions of Lorentz
indices. The replacements of \texttt{dEps} are consistent with
the dimension splitting in use and define the symbol \texttt{eps} such that
$n=4-2\varepsilon$. In the four-dimensional helicity (FDH) scheme~\cite{%
Bern:1991aq,Kunszt:1993sd,Catani:1996pk,Catani:2000ef,Signer:2008va}
the symbol \texttt{eps} can be replaced by zero; in the 't~Hooft-Veltman
scheme the products between $\varepsilon$ (resp.~$\varepsilon^2$) and the
form factors, which are formally Laurent series in $\varepsilon$, lead to
rational terms. After these steps the diagram is in a form suitable for
numerical evaluation if one provides routines for the computation of
spinor brackets ($\Spaa{\cdot\cdot}$ and $\Spbb{\cdot\cdot}$)
and a library of integral form factors such as
\texttt{golem95}~\cite{Binoth:2008uq}.

\begin{spform}
#call tHooftAlgebra
#call SpCollect
#call SpContract
#call SpContractMetrics
Id dEps(Q?{k1,k2,l3,l4}, iv1?) = 0;
Id dEps(iv1?, iv1?) = -2*eps;
#call SpOpen
\end{spform}

\subsubsection{Reduction at the Integrand Level}
Following the strategy of reducing the Feynman diagrams
at the integrand level, we show in this section how the numerator
of the Feynman diagram can be constructed very easily by the use of~\spinney{}.
We consider two different approaches corresponding to two publicly
available reduction packages. The strategy implemented in
\texttt{CutTools}~\cite{Ossola:2007ax} uses a numerator function $N(Q_4)$
which only depends on the four-dimensional projection of the (complex)
integration momentum $Q$. This method allows for the reconstruction of
the cut-constructible terms but only partially recovers the rational
parts of an amplitude. This is due to the fact that terms in $\mu^2$,
where $Q^2=Q_4^2-\mu^2$, lead to rational terms which are not taken into
account and need to be added by a separate
calculation~\cite{Draggiotis:2009yb,Garzelli:2009is}.
An improved reduction at the integrand level has been implemented by
the authors of \texttt{Samurai}~\cite{Mastrolia:2010nb}. This method takes advantage of
a numerator function depending on both $Q_4$ and $\mu^2$,
which can be decomposed into
\begin{equation}
N(Q_4,\mu^2)=N_0(Q_4,\mu^2)+\varepsilon N_1(Q_4,\mu^2)
   +\varepsilon^2 N_2(Q_4,\mu^2) + \mathcal O(\varepsilon^3).
\end{equation}
As in the case of the conventional tensor reduction,
$\varepsilon$~is set to zero (keeping terms in $\mu^2$) in the FDH scheme,
whereas in the 't-Hooft~Veltman scheme the independent reduction of the
numerators $N_0$, $N_1$ and $N_2$, multiplication with the
apropriate terms of the Laurent series of the scalar integrals
leads to the full result, including both both the cut-constructible
and the rational part of the amplitude. 

Irrespective of the approach chosen the first step to be taken should be
to carry out the 't~Hooft algebra. Then the two methods differ in the
way they deal with terms in \texttt{dEps}. Using \texttt{CutTools}
these terms are neglected and set to zero. In the case of \texttt{Samurai}
one substitutes the relations which are implied by $Q^2=Q_4^2-\mu^2$,
where $g^{\mu\nu}Q^2=Q_\mu Q_\nu$ and
$Q_4^\mu=\hat{g}^{\mu\nu} Q_\nu$, and therefore
$\mu^2=-\tilde{g}^{\mu\nu} Q_\mu Q_\nu$.
\begin{spform}
#call tHooftAlgebra
#call SpCollect
#call SpContractMetrics
Id dEps(Q,Q) = - mu2;
Id dEps(Q?, iv1?) = 0;
Id dEps(iv1?, iv1?) = -2*eps;
Argument Spaa,Spab,Spba,Spbb, d4;
   Id Q = Q4;
EndArgument;
Id Q.Q = Q4.Q4 - mu2;
Id Q = Q4;
Id d4(k1?, iv1?) = k1(iv1);
\end{spform}
Finally, one can contract the remaining Lorentz indices
and bring the the expression into a form where
spinor brackets ($\Spaa{\cdot\cdot}$, $\Spbb{\cdot\cdot}$ and
$\Spba{\cdot\vert\;/\!\!\!\!Q_4\vert\cdot}$) are the only functions
to be evaluated numerically.
\begin{spform}
#call SpContract
#call SpOpen(k1,k2,l3,l4)
\end{spform}

\subsection{Working with Majorana Spinors}
\subsubsection{Charge Conjugation of a Vector Current}
We illustrate our approach through the following simple example.
The vector current, $\langle p^{+} | \gamma^{\mu} | q^{+} \rangle$, satisfies the charge conjugation relation
\begin{equation}
\langle p^{+} | \gamma^{\mu} | q^{+} \rangle = 
\langle q^{-} | \gamma^{\mu} | p^{-} \rangle,
\end{equation}
which we can show explicitly using charge conjugation relations.
This charge conjugation operation is equivalent to us
reversing the fermion flow arrow. Therefore we can use Equations~\eqref{eq:fliprules} to show this equivalence
holds:
\begin{align}
\langle p^{+} | \gamma^{\mu} | q^{+} \rangle 
&= \brb{p} \mu \kea{q} \rightarrow (-1)^{P} \bra{q} \gamma^{\mu'} \keb{p} = (-1)^{P} (-1) \bra{q} \gamma^{\mu} \keb{p} \\
& = \bra{q} \gamma^{\mu} \keb{p} = \langle q^{-} | \gamma^{\mu} | p^{-} \rangle
\end{align}
where we have used \eqref{eq:etaproperties} and \eqref{eq:spinorCtransform} and we have defined our reference order 
as $(p,q)$ giving $(-1)^{P} = -1$.
This result is true for both Dirac and Majorana fermions.

\subsubsection{Majorana Exchange}

\begin{figure}[tbph]
\begin{center}$
\begin{array}{ccc}
\includegraphics[scale=0.45]{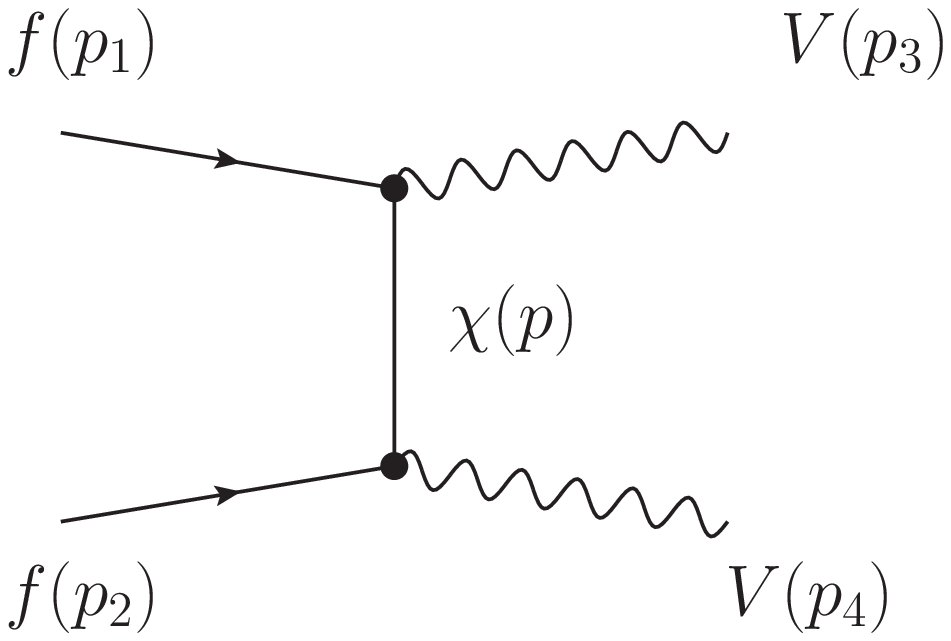}  &
\includegraphics[scale=0.45]{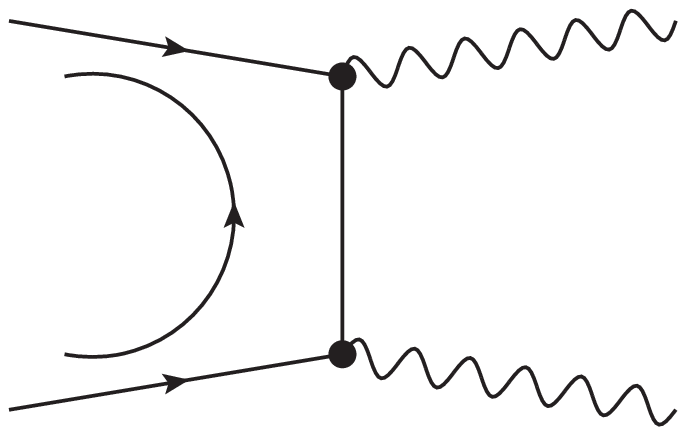} &
\includegraphics[scale=0.45]{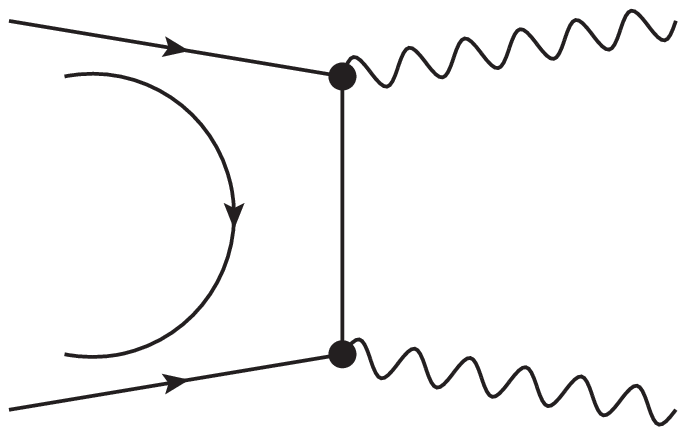}
\end{array}$
\end{center}
\caption{The process ($ff \rightarrow VV$) mediated by the exchange of a Majorana fermion ($\chi$) as
   discussed in the example. The left hand figure shows the original diagram; the middle figure 
shows the first choice of fermion orientation; and the figure on the right shows the 
second choice of orientation.}
\label{fig:ffVV}
\end{figure}

We now consider two Dirac fermions scattering to two vector
bosons via a t-channel Majorana fermion exchange. 
There are two equivalent orientations to choose from as shown in Figure~\ref{fig:ffVV}.
We write our amplitude as 
\begin{equation} 
\mathcal{A} = \mathcal{A}^{\mu \nu} \epsilon^{*}_{\mu}(p_{3})
\epsilon^{*}_{\mu}(p_{4}).
\end{equation}
Our first choice of orientation gives 
\begin{equation} \label{eq:ffVV1}
\mathcal{A}^{\mu\nu}_{1} = \langle p_{1} | (\gamma^{\mu})^{'} S(p) \gamma^{\nu} | p_{2} \rangle
\end{equation}
and the second orientation gives
\begin{equation}
\mathcal{A}^{\mu\nu}_{2} = (-1) \langle p_{2} | (\gamma^{\nu})^{'} S(-p) \gamma^{\mu} | p_{1} \rangle
\end{equation}
where we have chosen the reference order $(p_{1}, p_{2})$.
Applying our flipping rules to Equation~\eqref{eq:ffVV1} gives the result:
\begin{equation}
\mathcal{A}^{\mu\nu '}_{1} = (-1)^{P} \langle p_{2} | (\gamma^{\nu})^{'} S(-p) \gamma^{\mu} | p_{1} \rangle =
\mathcal{A}^{\mu\nu}_{2}.
\end{equation}
Therefore the amplitude is independent of the original choice of fermion orientation.

In our code, part of the output is:
\begin{form}
NCContainer(USpa(k1)*SpFlip(Sm(mu)*ProjPlus)*Sm(k4)
 *Sm(nu)*ProjPlus*USpb(k2))*inv(es23).
\end{form}
Upon applying the \verb|RemoveNCContainer| routine we obtain the result
\begin{form}   
- UbarSpb(k1)*ProjPlus*Sm(mu)*Sm(k4)*Sm(nu)
 *ProjMinus*USpa(k2)*inv(es23)
\end{form}
We have picked up a minus sign from the flipping of the $\gamma^{\mu}$. 
What remains is to multiply by $(-1)^{P}$ as explained previously.

\subsection{Coefficients of Scalar Integrals by Unitarity Based Methods}
\label{ssec:ex:unitarity}
In this example we consider Bhaba scattering in QED at the one-loop level.
We use the well known fact that any (leg-ordered) one-loop amplitude
can be written in terms of a basis of scalar one-loop integrals,
\begin{multline}
\mathcal{A}_{\text{1-loop}}=
\sum_{i_1} a_{i_1}\int\frac{\mathrm{d}^nl}{D_{i_1}}
+\sum_{i_1<i_2} b_{i_1i_2}\int\frac{\mathrm{d}^nl}{D_{i_1}D_{i_2}}
+\sum_{i_1<i_2<i_3} c_{i_1i_2i_3}%
    \int\frac{\mathrm{d}^nl}{D_{i_1}D_{i_2}D_{i_3}}\\
+\sum_{i_1<i_2<i_3<i_4} d_{i_1i_2i_3i_4}%
   \int\frac{\mathrm{d}^nl}{D_{i_1}D_{i_2}D_{i_3}D_{i_4}},
\end{multline}
where $D_{i}=[(l+r_i)^2-m_i^2]$.

\begin{figure}[h]
	\includegraphics[width=\textwidth]{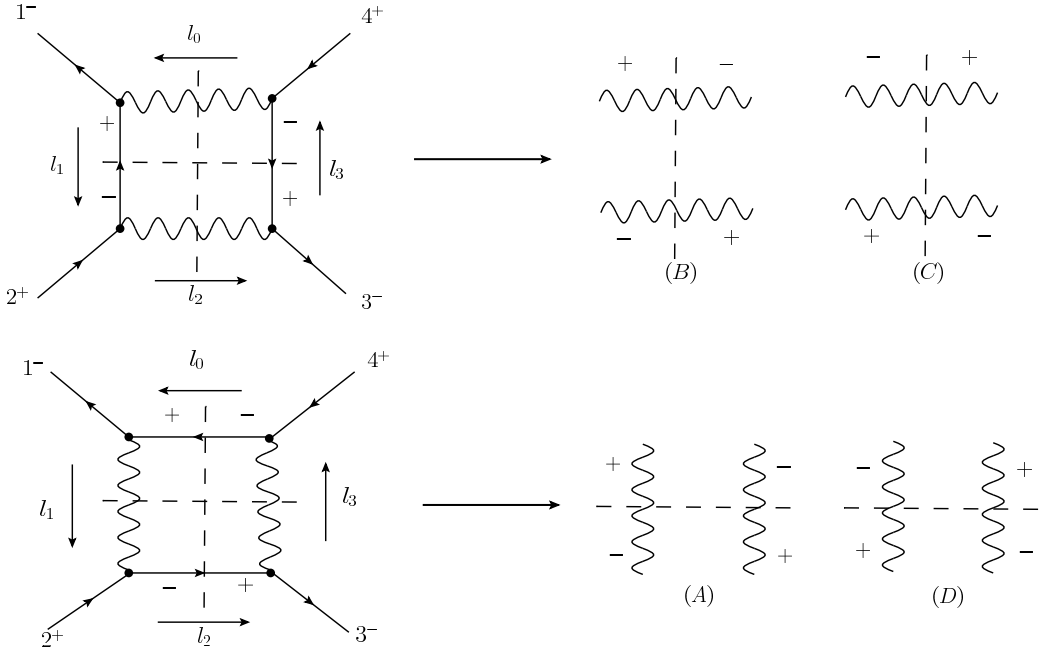}
	\caption{Possible helicity configurations of the two box diagrams
	contributing to the $f\bar{f}\rightarrow f\bar{f}$ amplitude.}
	\label{fig:box4cut}
\end{figure}

We consider the two diagrams in Fig.~\ref{fig:box4cut}.
The box coefficient is isolated by performing four cuts, 
which completely disconnect the loop amplitude into 4 tree-level
partial amplitudes with three external legs each.
In fact, since there is only one vertex in QED and the identity of one of
the legs is fixed (it is one of the original external fermions), 
we have two possible contributions in each amplitude differing by the 
exchange of the remaining fermion/photon legs of the QED vertex. 
Consistency, i.e. the fact that the lines connecting neighbouring vertices
need obviously be of the same type, reduces this to two choices,
depicted on the left-hand side of Fig.~\ref{fig:box4cut}.
The helicity conservation on the fermion lines and the restrictions of
complex kinematics allow us to write down the only four possible internal
helicity configurations.
The four cuts also provide four independent constraints on the loop momentum,
thus determining all its components (in fact there are two solutions).
For details see~\cite{Britto:2004nc}. We take all momenta to be outgoing.

We begin the computation by preparing a small procedure that will be used
repeatedly together with \texttt{SpOpen} to simplify denominator structures.
It is an ad-hoc solution based on the knowledge of type and depth of
denominators arising due to loop momenta substitutions.
\begin{spform}
#Include- spinney.hh
#Procedure SpSimplify()
   Repeat;
      FactArg SpDenominator;
      ChainOut SpDenominator;
      Id SpDenominator(SpDenominator(Spa2(vD1?,vD2?)))=
         Spa2(vD1,vD2);
      Id SpDenominator(SpDenominator(Spb2(vD1?,vD2?)))=
         Spb2(vD1,vD2);
      Id SpDenominator(SpDenominator(fD1?(?vD1)))= 
         fD1(?vD1); 
      Id SpDenominator(sDUMMY1?number_) = 1/sDUMMY1;
   EndRepeat;
#EndProcedure
\end{spform}
We write down the explicit formula for one of the four internal
helicity configurations, split into separate numerator and denominator
expressions. The formulae for the remaining configurations are similar.
\begin{spform}
Global FaNUM = -4 *Spbb(l1,l0) *Spaa(l0,l3) *Spaa(l1,l2) 
               *Spbb(l3,l2) *Spaa(p1,R0) *Spbb(R0,p2) 
               *Spaa(p3,R2) *Spbb(R2,p4) ;
Global FaDEN = Spaa(l1,R0) *Spbb(R0,l1) *Spaa(l3,R2) 
               *Spbb(R2,l3);
\end{spform}
The first step is to substitute the solutions for the loop momentum
on the cut, which in our case is:
\begin{equation}
l_0^{\mu} = \frac{1}{2}\frac{\Spbb{p2 p1}}{\Spbb{p2 p4}} 
            \Spab{p1\vert\gamma^{\mu}\vert p4}
\end{equation}
In order to do so we need to re-express all loop momenta in terms of $l_0$
and external momenta. We use \texttt{SpClose} to have the spinor products
in a form which is suitable for standard linear substitutions:
$\Spaa{p \vert \ldots l_{i_1} \ldots l_{i_2}\ldots \vert q}$, 
with $p$ and $q$ being external momenta and $l_i$ the loop momentum.
This might not be immediately possible,
as in our example, where \texttt{SpClose} produces an output of
the type $\Spaa{l_{i_1} \vert\ldots \vert l_{i_2}}$ instead.
However, a call to \texttt{Schouten} solves the problem at the cost of
having more terms in the calculation.
\begin{spform}
#Call SpOpen
#Call Schouten(l0,l3,p1,R0)
#Call SpClose(l0,l1,l2,l3)
.sort
ToTensor Functions l1, L1;
ToTensor Functions l2, L2;
ToTensor Functions l3, L3;
Id L1(iD1?) = l0(iD1) - p1(iD1);
Id L2(iD1?) = l0(iD1) - p1(iD1) - p2(iD1);
Id L3(iD1?) = l0(iD1) + p4(iD1);
#Call SpContract
.sort
\end{spform}
Now we substitute the solution for $l_0^{\mu}$ and contract all explicit
Lorentz~indices.
\begin{spform}
ToTensor Functions l0, L0;
ChainOut L0;
.sort
Id L0(iD1?) = Spab(p1,iD1,p4)*Spb2(p2,p1)*(1/2)*
              SpDenominator(Spb2(p2,p4));
#Call SpContractMetrics
#Call SpContract
#Call SpOpen
.sort
\end{spform}
We combine the numerator and denominator expressions obtained before.
Our formulae still involve the reference momenta of the photons,
which is our gauge freedom - we can greatly simplify the computations
by making an explicit choice at this point.
\begin{spform}
Global Fa = FaNUM*SpDenominator(FaDEN);
Argument;
   Id R2 = p1;
   Argument;
      Id R2 = p1;
   EndArgument;
EndArgument;
#Call SpOpen
#Call SpSimplify
#Call SpOpen
.sort
\end{spform}
As a finishing touch, we apply the following sequence of transformations
to get rid of any remaining double denominator-type expressions,
which appear upon introducing the explicit form of the $l_0$ loop~momentum.
\begin{spform}
Argument;
   #Call Schouten(p1,p2,p4,R0)
EndArgument;
Repeat;
   #Call SpOpen()
   Argument;
      #Call SpOpen()
   EndArgument;
   #Call SpSimplify
EndRepeat;
.store
\end{spform}

Finally, we write down the expression for the complete box coefficient.
Note the factor $1/2$, which comes from averaging over two solutions to the
loop momentum constraints.
\begin{spform}
Local F = (1/2)*(Fa + Fb + Fc + Fd);
Print F;
.end
\end{spform}

The \texttt{FORM} output we obtain is:
\begin{multline}\label{eq:UN3}     
F = -\frac{2\Spaa{p_1  p_2}^3\Spaa{p_3  p_4}\Spbb{p_2  p_1}^2}{
    \Spaa{p_2  p_4}^2} + \frac{2\Spaa{p_1  p_4}^3\Spaa{p_2  p_3}
    \Spbb{p_4  p_1}^2}{\Spaa{p_2  p_4}^2} \\
   +2\Spaa{p_1  p_2}\Spaa{p_1  p_3}\Spbb{p_2  p_1}\Spbb{p_4  p_2} - 
 2\Spaa{p_1  p_3}\Spaa{p_1  p_4}\Spbb{p_4  p_1}\Spbb{p_4  p_2}.
\end{multline}

\section{Conclusion}\label{sec:conclusion}
In this article we have presented a new \FORM{} library for processing
expressions containing helicity spinors in four and
$n=(4-2\varepsilon)$ dimensions.
For the $n$-dimensional algebra we have implemented the 't-Hooft algebra
with dimension splitting.
This gives full flexibility to the user about the choice of the
regularisation scheme, as in many schemes different from
the 't~Hooft-Veltman scheme it is sufficient to neglect terms
in $\varepsilon$ which stem from the numerator algebra.

In various examples we have shown that the new package is applicable
to calculations both using traditional and modern, unitarity based
methods. The provided routines implement the typical steps which are
necessary for an algebraic simplification of helicity amplitudes
and therefore simplify the task of implementing such calculations.
The implementation of flipping rules for Majorana spinors allows to
extend the domain of applicability to theories beyond the Standard Model.

\section*{Acknowledgements}
T.R. wants to thank Ralf Sattler for useful discussion.
T.R. is supported by the Dutch Foundation for Fundamental Research
on Matter~(FOM), project FORM~07PR2556.
GC would like to thank Nikhef for their hospitality while part of this 
work was carried out.
G.C. is supported by the British Science and Technology Facilities
Council~(STFC).


\bibliography{spinney-cpc}

\end{document}